\documentclass[aps,prx,twocolumn,floatfix,superscriptaddress,longbibliography]{revtex4-1}

	\usepackage{amsmath}
	\usepackage{amssymb}
	\usepackage{float}
	\usepackage{amsthm} 
	\usepackage{nicefrac}
	\usepackage{braket}
	\usepackage{enumerate}
	\usepackage[pdftex]{graphicx}
	\usepackage{txfonts}
	\usepackage{dcolumn} 
	\usepackage{color}
	\usepackage{bm}
	\usepackage[version=3]{mhchem}
	\usepackage[colorlinks,citecolor=blue,linkcolor=blue,urlcolor=blue]{hyperref}
	\newcommand{\abs}[1]{\left| #1 \right|} 
	\newcommand{\avg}[1]{\left< #1 \right>} 
	\newcommand{\trace}{\mbox{Tr}} 

\begin{document}
	\title{Phase Diagram of a Spin-ice Kondo Lattice Model in a Breathing Pyrochlore Lattice}
	\author{Munir Shahzad}
	\affiliation{Department of Physics and Physical Oceanography, Memorial University of Newfoundland, St.  John’s, Newfoundland \& Labrador A1B 3X7, Canada}
	\author{Kipton Barros}
	\affiliation{Theoretical Division and CNLS, Los Alamos National Laboratory, Los Alamos, NM 87545, USA}
	\author{Stephanie H. Curnoe}
	\affiliation{Department of Physics and Physical Oceanography, Memorial University of Newfoundland, St.  John’s, Newfoundland \& Labrador A1B 3X7, Canada}
	\date{\today}
	\begin{abstract}
	We study a spin-ice Kondo lattice model on a breathing pyrochlore lattice with classical localized spins. The highly efficient kernel polynomial expansion method, together with a classical Monte Carlo method, is employed in order to study the magnetic phase diagram at four representative values of the number density of itinerant electrons. We tune the breathing mode by varying the hopping ratio -- the ratio of hopping parameters for itinerant electrons along inequivalent paths. Several interesting magnetic phases are stabilized in the phase diagram parameterized by the hopping ratio, Kondo coupling, and electronic filling fraction, including an  ``all-in/all-out" ordered spin configuration phase, spin-ice, ordered phases containing $16$ and $32$ spin sites in the magnetic unit cell, as well as a disordered phase at small values of the hopping ratio. 

	\end{abstract}
	\maketitle

    \section{Introduction}\label{sec:intro}
    Conducting pyrochlore magnets
    \ce{R2B2O7} (\ce{R} = Pr, Nd, Sm or Eu, \ce{B} = Ir or Mo)
    present a dual challenge: 
    on the one hand, they are geometrically frustrated spin systems with corner-sharing tetrahedral networks 
    on both the \ce{R} and \ce{B} sites, 
    while on the other they are correlated metals.
    The rare earth spins \ce{R} 
    interact via direct exchange or via RKKY exchange
    originating from the Kondo interaction between
    the \ce{B} site conduction electrons and the
    \ce{R} site local moments.
    The \ce{B} site conduction bands can be described
    within tight-binding models, which may include spin-flipping terms
    brought about by spin-orbit coupling, and local (Hubbard)
    interactions.
    All of this takes place on
    a highly symmetric crystal which imposes strict constraints on the models \cite{curnoe2008, onada2011, sblee2013, eklee2013,  curnoe2013, huang2014}.         
    
    The interplay between the \ce{B} site conduction electrons and the \ce{R} site local moments in pyrochlore conductors
    effectuates a rich magnetic phase diagram.
    While \ce{Pr2Ir2O7} is metallic, the other
    rare-earth iridates (\ce{R} = Nd, Sm and Eu)
    undergo metal-insulator phase transitions
    at temperatures in the range of 36 to 120 K \cite{matsuhira2007}.
    Aside from the obvious differences in transport, these metals differ 
    from their insulating cousins in their magnetic properties.
    Magnetization measurements on
    the iridates are indicative of 
    antiferromagnetic (AFM) correlations between the rare earth sites,
    with possibly AFM ordering of Nd spins \cite{yanagishima2001}. 
    This stands in contrast to their ruthenium analogs
    \ce{R2Ru2O7} 
    which are spin glasses or weak ferromagnets \cite{taira1999}.
    Generally these effects can be attributed to the the Kondo interaction between conduction electrons and local spins which augments the exchange interaction between the rare earth spins (in so-called ``double-exchange" models \cite{ikoma-magnetic2003}) or induces effective RKKY magnetic interactions between
    the rare earth spins \cite{ikeda_ordering_2008, flint2013,sblee2013}. In fact, a minimum of the resistivity, the hallmark feature of the Kondo effect, has been observed in \ce{Pr2Ir2O7}
    \cite{nakatsuji2006}.
    Also, Ir clearly plays a role inducing magnetic interactions in Lu and Y iridates since Lu and Y are otherwise non-magnetic; moreover,
    while the spin ice compounds 
    \ce{Ho2Ti2O7} and \ce{Dy2Ti2O7} have ferromagnetic (FM) interactions,
    their (insulating) iridate cousins are antiferromagnetic \cite{yanagishima2001}.
    Similarly, Mo plays an important role in the conducting mobdylates  \ce{R2Mo2O7} (\ce{R}= Nd, Sm, Gd).
    A large anomalous Hall effect in Nd$_2$Mo$_2$O$_7$ (a conducting ferromagnet) is attributed to a chiral spin arrangement of (predominantly) Mo spins resulting from their coupling to Nd moments \cite{taguchi_spin_2001}.
    A chiral spin configuration also occurs in \ce{Pr2Ir2O7} due to a non-coplanar 
    arrangement of Pr spins \cite{machida_unconventional_2007}.
    
    Generally frustration tends to impede long-range order
    in systems with AFM correlations, but
    frustration effects can be reduced by structural changes. Cubic to tetragonal lattice
    distortions accompanying magnetic order in  spinel oxides  \ce{AB2O4} (\ce{A}= Mg, Cd, Zn, \ce{B}= Cr, V)  \cite{mamiya-structural1997,ueda-magnetic1996,lee_local2000, chung2005, ueda2005},
    as well as \ce{ZnCr2Se4} \cite{hemberger-large2007},
    are
    well-documented.
    These distortions are associated with lifting of the spin degeneracy (due to frustration) via magneto-elastic interactions \cite{yamashita2000, tchernyshyov-order2002,tchernyshyov-spin2002, chern-broken2006}. More recently, there has been a heightened interest in the breathing pyrochlore and kagome lattices \cite{okamoto_breathing_2013, kimura_experimental_2014, okamoto_magnetic_2015, haku_low-energy_2016, rau_anisotropic_2016, okamoto_breathing_2018, pokharel_negative_2018,  aoyama-spin2019}. A breathing lattice consists of alternating large and small neighboring units -- tetrahedra and triangles for pyrochlore (or spinel) and kagome lattices, respectively. These lattices have been realized experimentally~\cite{orain_2017_nature,okamoto_breathing_2013,kimura_experimental_2014,okamoto_magnetic_2015,haku_low-energy_2016} and exhibit interesting phenomena such as helical and skyrmion magnetic phases~\cite{hirschberger_skyrmion_2019,khanh_nanometric_2020,hirschberger_high_2020,takuya_formation_2020,Cameron_2016}, the existence of a Weyl magnon (the  bosonic analogue of a Weyl fermion)~\cite{ezawa_higher-order_2018,li_weyl_2016,jian_weyl_2018}, and negative thermal expansion as a result of strong magnetoelastic coupling~\cite{hemberger_large_2007}.
    
   In pyrochlore and spinel crystals the breathing mode does not change the crystal system -- it remains cubic -- but it does remove some of the point group symmetry elements,
    resulting in a lowering of the 
    space group symmetry.
    The \ce{R} and \ce{B} sites on the pyrochlores and spinels form a corner-sharing tetrahedral lattice on which the tetrahedra alternate between two orientations. The breathing mode amounts to one orientation of tetrahedra expanding while the other contracts, with a change in the space group symmetry from $Fd{\bar 3}m$ to $F\bar{4}3m$.
    The alternation in size between neighboring units results in different inter-atomic interaction strengths along paths within to each neighboring unit, introducing a concomitant inequality in the exchange constants \cite{curnoe2008} and hopping parameters between the alternating
    tetrahedra. The tetrahedra are completely decoupled in the limit where these parameters vanish on one set of tetrahedra.
    
    The \ce{Cr}-based spinels, such as \ce{Li, (In, Ga)Cr4O8} and \ce{Li, (In, Ga)Cr4S8}, are breathing lattices in which the relative size difference between the neighboring tetrahedra is small (between $1.05$ and $1.1$)~\cite{okamoto_breathing_2013,okamoto_magnetic_2015,okamoto_breathing_2018,pokharel_negative_2018}. These compounds represent the `strongly coupled' limit and there is a transition to a magnetic ground state in most of these compounds. On the other hand, the compound \ce{Ba3Yb2Zn5O11} is in the opposite limit where the modulation in size between neighboring tetrahedra is $\sim 2$~\cite{kimura_experimental_2014,haku_low-energy_2016,rau_anisotropic_2016}. The residual entropy and absence of magnetic order in this compound can be attributed to decoupled tetrahedra. The investigation and hence modelling of the compounds between these extreme limits is thus a timely enterprise.

In this work, we numerically study the magnetic phase diagram of a spin-ice Kondo lattice model in a breathing pyrochlore lattice using a kernel polynomial expansion method together with an unbiased classical Monte Carlo method. 
    The breathing mode is incorporated in terms of a ratio of hopping amplitudes (the `hopping ratio') of itinerant electrons on alternating tetrahedra on the pyrochlore lattice. A related study has been done on an isotropic lattice at small Kondo coupling~\cite{ishizuka_magnetic_2012,ishizuka_application_2012,ishizuka_polynomial_2013}. In the present work, we not only include large Kondo coupling but also study the effects of the breathing mode on the magnetic phase diagram. Our study reveals the existence of several interesting phases including an all-in/all-out (AIAO) spin configuration (an ordered arrangement in which one orientation of the tetrahedra has all four spins pointing in towards the
    centres of the tetrahedra, while the other orientation has the four spins pointing out from the centres), a
  spin-ice (SI) phase (a disordered arrangement in which two spins point into and two spins point out of each tetrahedron), and ordered phases in which the magnetic unit cell consists of 16 sites or 32 sites are stabilized over wide ranges of Kondo coupling and hopping ratio.
		
	\section{Model}\label{sec:model}
\begin{figure}[thb]
	\centering
	\includegraphics[clip,trim=0cm 0cm 0cm 0cm,width=0.35\textwidth]{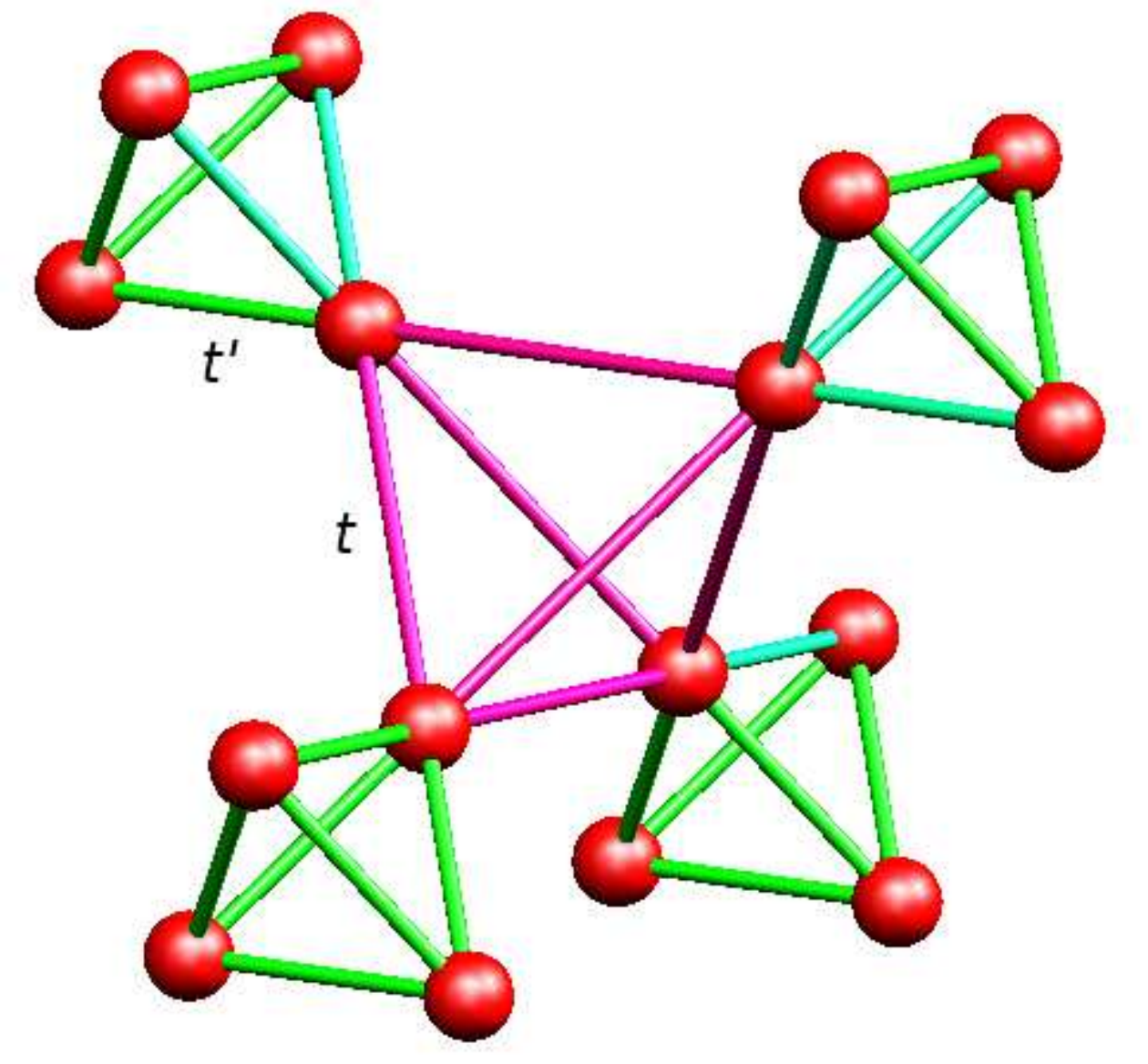}
	\caption{(Color online) 
	The corner-sharing tetrahedral network of 
	pyrochlore and spinel crystals. 
	The tetrahedra occur in two 
	different orientations, up-pointing (green)
	and down-pointing (pink). In a breathing lattice the hopping constants $t$ and $t'$ associated with the edges of each kind of tetrahedron are not equal.
	\label{fig:tetrahedra}}
	\end{figure}

	We investigate the magnetic properties of localized spins in a Kondo lattice model on a breathing pyrochlore lattice. The Hamiltonian for a spin-ice Kondo lattice model on a breathing pyrochlore lattice can be written as
	\begin{multline}
	\label{equ:ham}
	\mathcal{\hat{H}}= -t\sum_{\avg{i,j}\in d,\sigma}(c_{i\sigma}^\dagger c_{j\sigma}+\mathit{H.c.})-t'\sum_{\avg{i,j}\in u,\sigma}(c_{i\sigma}^\dagger c_{j\sigma}+\mathit{H.c.})\\-J_K \sum_i \mathbf{S}_i\cdot \mathbf{s}_i,
	\end{multline}					
	where $t$ and $t'$ are nearest neighbor hopping amplitudes on down-pointing and up-pointing tetrahedra, respectively (see Fig.~\ref{fig:tetrahedra}) and $J_K$ is the strength of the on-site Kondo interaction between the localized spins $\mathbf{S}_i$ and the spins of conduction electrons $\mathbf{s}_i$. We assume the localized spins to be Ising spins with $\abs{\mathbf{S}_i}=1$ and the anisotropy axes of these spins are their local three-fold symmetry axes, {\em i.e.}, the $\avg{111}$ direction. This direction is parallel to the line connecting the centers of the two neighboring tetrahedra to which spin belongs. With the help of Pauli matrices, the spin of the conduction electron can be written in terms of raising and lowering operators as $\mathbf{s}_i= c_{i\alpha}^\dagger\bm{\sigma}_{\alpha\beta}c_{i\beta}$. In the present model, the sign of $J_K$ (ferromagnetic or antiferromagnetic) is irrelevant as the eigenstates that correspond to different signs of $J_K$ are related by a global gauge transformation~\cite{PhysRevB.72.075118,PhysRevLett.101.156402}. The hopping ratio is $t'/t$, and from here onwards, we take the hopping amplitude $t=1$ as the energy unit.
	
	\section{Method and observables}\label{sec:methods}
				
	To investigate the above model, we use two methods, the exact diagonalization and Monte Carlo method (ED-MC) and the kernel polynomial expansion and Monte Carlo method (KPM-MC). The fundamental difference between these two methods is the way they evaluate the trace over fermionic degrees of freedom. The dynamics of large localized moments is slow compared to itinerant electrons, and accordingly, we can decouple their dynamics from that of the itinerant electrons. Effectively, we treat the local moments as classical fields at each site. The Hamiltonian in Eq.~\eqref{equ:ham} is bilinear in fermionic operators and can be represented as
	\begin{equation}
	\label{equ:bilinear_ham}
	\mathcal{\hat{H}}= \sum_{i,j}c_{i\sigma}^\dagger H_{ij}(\{\phi_r\}) c_{j\sigma}.
	\end{equation}
	In the single-electron basis, $H_{ij}(\{\phi_r\})$ is a $2\mathit{N}\times 2\mathit{N}$ matrix for a fixed configuration of classical localized Ising spins $\phi_r$, where $\mathit{N}$ is the number of sites. 
					
	In order to explore the thermodynamic properties, we write the partition function for the whole system by taking two traces,	
	\begin{equation}
	\label{equ:full_trace}
	Z=\trace_c\trace_f \exp({-\beta[\mathcal{\hat{H}}({\{\phi_r\}})-\mu \hat{n}_e]}),
	\end{equation}
	where $\trace_c$ and $\trace_f$ are the traces over the classical localized spins and the itinerant electron degrees of freedom, respectively. The trace over itinerant electron degrees of freedom is calculated by one of two methods, exact diagonalization or kernel polynomial expansion method (KPM). In the first method, a numerical diagonalization of the Hamiltonian matrix $H(\{\phi_r\})$ is performed in order to evaluate the $\trace_f$ 
	using the eigenvalues $\varepsilon_\nu(\{\phi_r\})$:
	\begin{multline}
	\label{equ:trace}
	\trace_f \exp\{{-\beta[\mathcal{\hat{H}}}(\{\phi_r\})-\mu \hat{n}_e]\}\\
	\equiv\prod_\nu(1+\exp\{-\beta[\varepsilon_\nu(\{\phi_r\})-\mu]\}),
	\end{multline}
	where $\mu$ is the chemical potential, $\beta=1/k_BT$ is the inverse temperature, and $\hat{n}_e=\frac{1}{2N}\sum_{i\sigma}c_{i\sigma}^\dagger c_{i\sigma}$ is the number density operator for conduction electrons. The partition function for the whole system then takes the form
	\begin{equation}
	\label{equ:part_func}
	\mathit{Z} =\trace_c \exp[-S_{\mathit{eff}}(\{\phi_r\})].
	\end{equation}
	The corresponding effective action is
	$S_{\mathit{eff}}(\{\phi_r\})=\sum_\nu F(\varepsilon_\nu(\{\phi_r\}))$, where $F(y)=-\ln[1+\exp\{-\beta(y-\mu)\}]$. A disadvantage of this approach is that direct diagonalization of the single-particle Hamiltonian matrix $H$ has a numerical cost that scales cubically in system size $N$.
	
	To speed up the calculations of $S_{\mathit{eff}}(\{\phi_r\})$, we make use of the KPM~\cite{Silver94, Weisse06}. The key idea in KPM is to write $S_{\mathit{eff}}(\{\phi_r\})=\mathrm{Tr}\, F(H)$, and then to expand $F[H]$ in Chebyshev matrix polynomials up to some fixed order $M$. The appropriate cutoff $M$ will typically need to be larger at lower temperatures, which allows for finer resolution of the density of states near the Fermi surface. If one additionally employs a stochastic approximation of the trace, $\mathrm{Tr}\, F(H) \approx \mathrm{Tr}\, R^\dagger F(H) R$, where $R$ is a suitable random matrix, the computational cost scales linearly with system size, assuming $H$ is sparse. For this study, we use the deterministic variant of KPM, for which the computational cost scales quadratically in system size. See Appendix~\ref{sec:KPM} for more details of the method. Our implementation of KPM uses the Nvidia CuSPARSE library for highly efficient execution on graphical processing unit (GPU) hardware.
	
	The grand-canonical trace over localized spin degrees of freedom in Eq.~\eqref{equ:full_trace} is evaluated by sampling the spin configuration space using a Monte Carlo (MC) method. The probability distribution for a particular configuration of localized spins $\{\phi_r\}$ can be written as
	\begin{equation}
	\label{equ:prob_dist}
	\mathit{P(\{\phi_r\})}\propto \exp[-\mathit{S_{eff}}(\{\phi_r\})].
	\end{equation}
	The thermodynamic quantities that depend on localized spins are calculated by the thermal averages of spin configurations, while the quantities that are associated with itinerant electrons are calculated from the eigenvalues and eigenfunctions of $H(\{\phi_r\})$. We start the simulations with a random configuration of Ising spins $\{\phi_r\}$ and calculate the Boltzmann action
	$S_{\mathit{eff}}(\{\phi_r\})$ for this configuration. The spin configuration is updated via the Metropolis algorithm based on the change in the effective action resulting from random single spin flip updates, $\Delta S_{\mathit{eff}}=S_{\mathit{eff}}(\{\phi_r'\})-S_{\mathit{eff}}(\{\phi_r\})$. Because the spin degrees of freedom are discrete, we cannot use a continuous Langevin dynamics to sample $\phi_r$, as in previous work~\cite{barros_efficient_2013,chern_semiclassical_2018}. 

	To identify different magnetic orderings we calculate the order parameter $P^{\alpha}_\mathbf{q}$ defined as
	\begin{equation}
	\label{equ:p_q}
	P^{\alpha}_\mathbf{q} = \frac{\max\left[S^{\alpha}(\mathbf{q})\right]}{N_t},
	\end{equation}
	where $\max\left[S^{\alpha}(\mathbf{q})\right]$ is the magnitude of the highest peak in the sublattice spin structure factor $S^{\alpha}(\mathbf{q})$, which is the Fourier transform of the spin-spin correlation function,		\begin{equation}
	\label{equ:str_fact}
	S^{\alpha}(\mathbf{q})=\frac{1}{N_t}\sum_{i,j\in \alpha }\avg {\mathbf{S}_i \cdot \mathbf{S}_j} \exp [i\mathbf{q} \cdot \mathbf{r}_{ij}].
	\end{equation}
	In the above equation, $\alpha = A, B, C, D$ denotes the $4$ inequivalent sub-lattices inside a primitive unit cell of the pyrochlore lattice and $\mathbf{r}_{ij}$ is the position vector from the $\mathit{i}$th site to the $\mathit{j}$th site. The sum is over nearest neighbours at sites $i$ and $j$, where $j$ is a type $\alpha$ site. $N_t=N/4$ is the total number of tetrahedra, and $\avg{\cdot}$ represents the thermal average over the grand-canonical ensemble. Additionally, we examine local spin correlations by calculating the fraction of tetrahedra with all-in or all-out ($P_{40}$), 3-in-1-out or 3-out-1-in ($P_{31}$) and 2-in-2-out ($P_{22}$) spin configurations. $P_{40} = 1$ in the AIAO phase, $P_{22} = 1$ in the SI phase, and for a completely random configuration $P_{40} = 2/16$,
	$P_{31} = 8/16$ and $P_{22} = 6/16$.

	\section{Results and Discussion}\label{sec:results}

	The numerical methods described in the previous section are used to perform the simulations of the model \eqref{equ:ham} for lattices sizes of $N_t=4^3$ to $8^3$ over a range of the Hamiltonian parameters $J_k$ and $t'/t$. All results reported below were calculated using the KPM-MC method with polynomial expansion order $M=1000$. We selected this $M$ value by validating against ED-MC simulations for small system sizes. We use the simulated annealing method to prevent freezing of the local moments that may occur at low temperatures. In this method, we generally start the simulation with a spin configuration at a comparatively high temperature ($T=2.0$ in this case) and perform MC equilibration steps in order to find the minimum energy configuration at that temperature.
	Next, we decrease the temperature by $\Delta T$ and use the final spin configuration from the previous $T$ as the initial configuration for the new value of the temperature. We repeat this process until we reach $T=0.001$, at which point measurements are performed in order to calculate the thermal averages of physical observables. We used $30$ temperature steps and a total of $60,000$ MC steps for equilibration, and a further $2,000$ steps were used to perform the measurements of the observables. 
	
	The Hamiltonian in~\eqref{equ:ham} is presumed to have a rich phase diagram owing to a large number of parameters involved. In the present work, we determine the magnetic phase diagram at four representative values of the number density of itinerant electrons, $n_e=1/2$, $1/3$, $1/4$ and $1/6$
	(where $n_e = \avg{\hat{n}_e}$), while varying the hopping ratio $t'/t$ and the Kondo coupling $J_K$. We benchmarked our results with previously published results for $J_K=2$ on an isotropic pyrochlore lattice and realized all of the magnetic phases in the phase diagram presented therein~\cite{ishizuka_magnetic_2012,ishizuka_application_2012,ishizuka_polynomial_2013}.
	\begin{figure}[htb]
	\centering
	\includegraphics[clip,trim=0cm 0cm 0cm 0cm,width=0.40\textwidth]{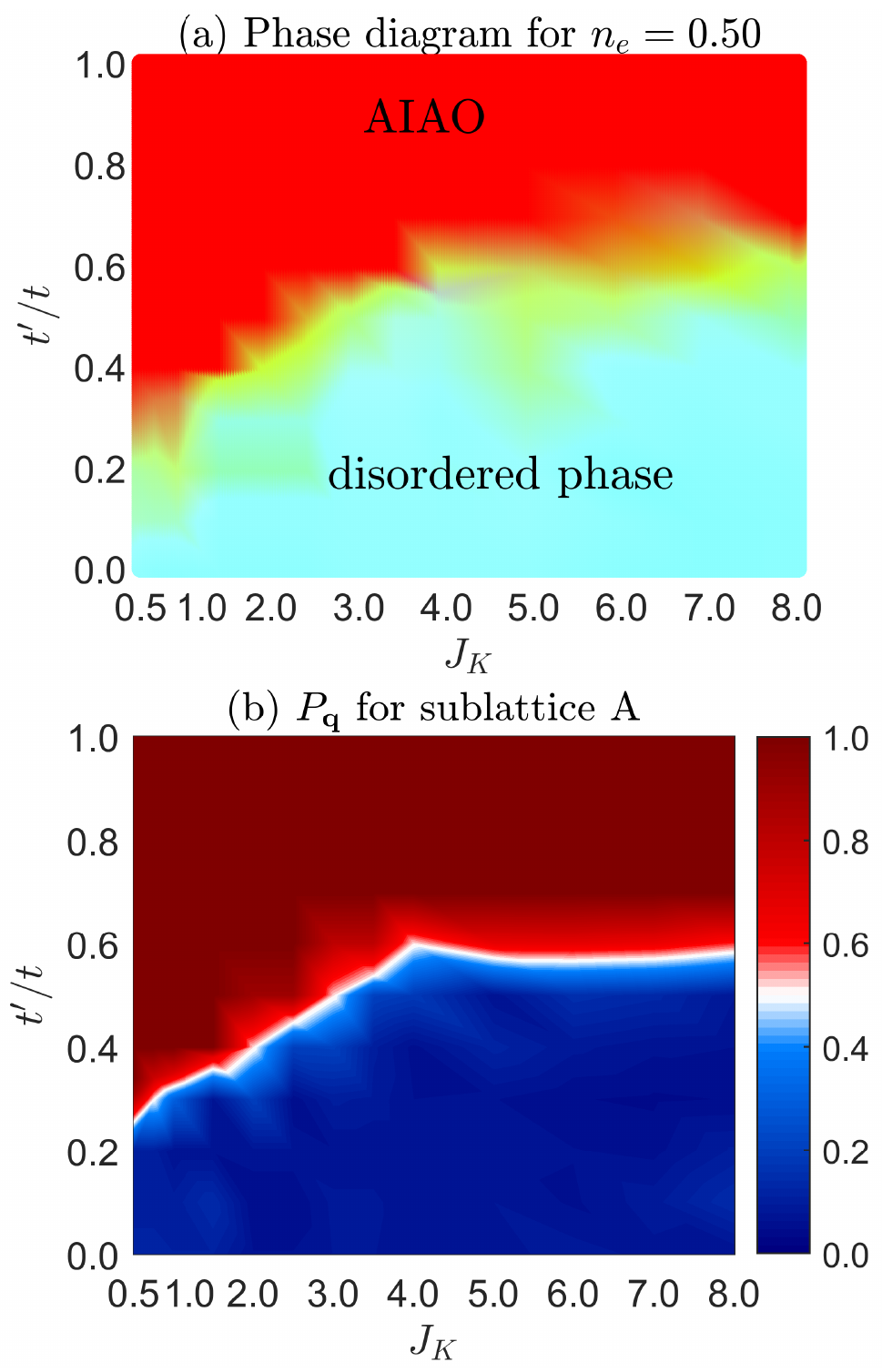}
	\caption{(Color online) (a) The phase diagram for $n_e=0.50$ as a function of Kondo coupling $J_K$ and hopping ratio $t'/t$. A RGB color scheme is used to draw the diagram, where the local correlation fractions $P_{40}$, $P_{31}$ and $P_{22}$  are expressed in terms of red, green and blue colors, respectively. 
	(b) $P_\mathbf{q}$ for sublattice $A$ plotted as a function of $J_K$ and $t'/t$.  The boundary between the ordered and disordered phases is clearly evident.}
	\label{fig:phase_diag_mq_ne_0.50}
	\end{figure}
	\subsection{One-half filling}\label{subsec:half_filling}
	We start our discussion by analyzing the phase diagram of localized spins when number density of itinerant electrons is 0.5. The evolution of the magnetic ground state as a function of $t'/t$ and $J_K$ is shown in Fig.~\ref{fig:phase_diag_mq_ne_0.50}(a), where we have represented the local correlation fractions $P_{40}$, $P_{31}$ and $P_{22}$ with weighted mixtures of red, green and  blue colors, respectively. 
	There are two magnetic phases present in the phase diagram, an AIAO phase and a disordered phase. For the isotropic pyrochlore lattice (where $t'/t=1.0$), for all values of Kondo coupling, we realized an AIAO ground state. 
	
	At large $J_K$ one expects the double exchange mechanism to govern the phase diagram. The fermionic kinetic energy (K.E.) stabilizes FM ordering of the localized spins as there is large K.E. gain if the spins on two neighbouring sites are parallel. However, at half filling of the itinerant electrons this argument is not valid as the lower bands are completely filled and an energy of the order of $J_K$ is required to cause the hopping hence AFM ordering of the localized spins is favored. For a pyrochlore lattice with Ising spins, AFM correlations are  not frustrated, rather they stabilize the AIAO ordered state. Similarly, for the isotropic pyrochlore lattice, at small values of $J_K$, a second order perturbation in terms of $J_K/t$ results in an effective RKKY Hamiltonian as shown in Ref.~\onlinecite{ishizuka_magnetic_2012}. Therefore, for $n_e=0.50$, the dominant nearest neighbor component of the RKKY interaction is AFM and AIAO order is stabilized.
	
	With a decrease of the hopping ratio $t'/t$, at both strong and weak $J_K$ coupling, $P_{40}$, the fraction of all-in or all-out tetrahedra, decreases and a phase transition is observed where the ground state changes from the ordered AIAO phase to a disordered phase with predominantly all-in or all-out configurations. The phase transition between the ordered and disordered phases is also evident when we consider the order parameter $P_\mathbf{q}$ for sublattice $A$
	shown in Fig.~\ref{fig:phase_diag_mq_ne_0.50}(b). The ordered AIAO phase is manifested as a sharp peak in the spin structure factor at $\mathbf{q}=(0,0,0)$ for all four sublattices. The magnitude of the peak decreases as  the hopping ratio $t'/t$ decreases, for small and large Kondo coupling. At the bottom of the phase diagram the magnitude of the peak in $S(\mathbf{q})$ (and hence $P_\mathbf{q}$) is small, indicating a disordered phase.
	\begin{figure}[tb]
	\centering
	\includegraphics[clip,trim=0cm 0cm 0cm 0cm,width=0.49\textwidth]{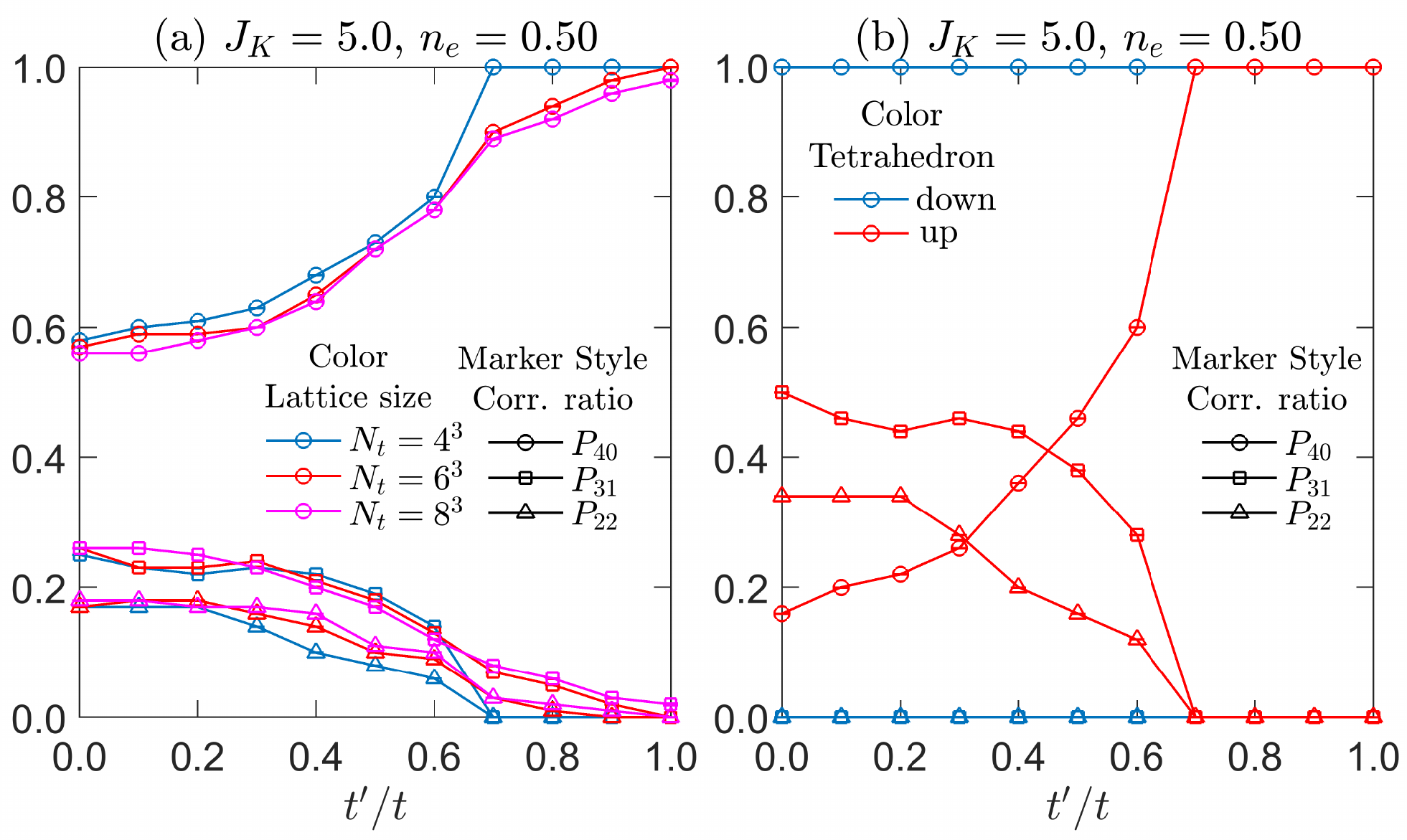}
	\caption{(Color online) (a) Local correlation fractions $P_{40}$, $P_{31}$ and $P_{22}$ vs. hopping ratio $t'/t$ for $J_K=5.0$  and system sizes $N_t = 4^3$, $6^3$ and $8^3$. (b) Local correlation fractions as a function of $t'/t$ at $J_K=5.0$ for up-pointing 
	and down-pointing tetrahedra and system size $N_t=4^3$.}
	\label{fig:line_plots_ne_0.50}
	\end{figure}
	
	We also plot the local correlation fractions for different system sizes in Fig.~\ref{fig:line_plots_ne_0.50}(a) as a function of $t'/t$ for $J_K=5.0$. The results for different lattice sizes are consistent with each other and show a transition from an ordered AIAO state to a disordered state when the hopping ratio is varied. For the model we considered, there is no direct exchange interaction between localized spins; instead interactions are mediated by the itinerant electrons hopping from site to site. When the hopping ratio $t'/t$ is close to one there is an AIAO ordered state; as $t'/t$ decreases  
	the state changes to one with all-in or all-out configurations on all down-pointing tetrahedra (half of the down-pointing tetrahedra have all-in while other half have all-out pointing spins) and to a state with all types of spin configurations on the up-pointing tetrahedra. 
	We show this effect in  Fig.~\ref{fig:line_plots_ne_0.50}(b), where the local correlations fractions are plotted for up- and down-pointing tetrahedra as a function of $t'/t$ and at $J_K=5.0$. 
	\begin{figure}[htb]
	\centering
	\includegraphics[clip,trim=0cm 0cm 0cm 0cm,width=0.40\textwidth]{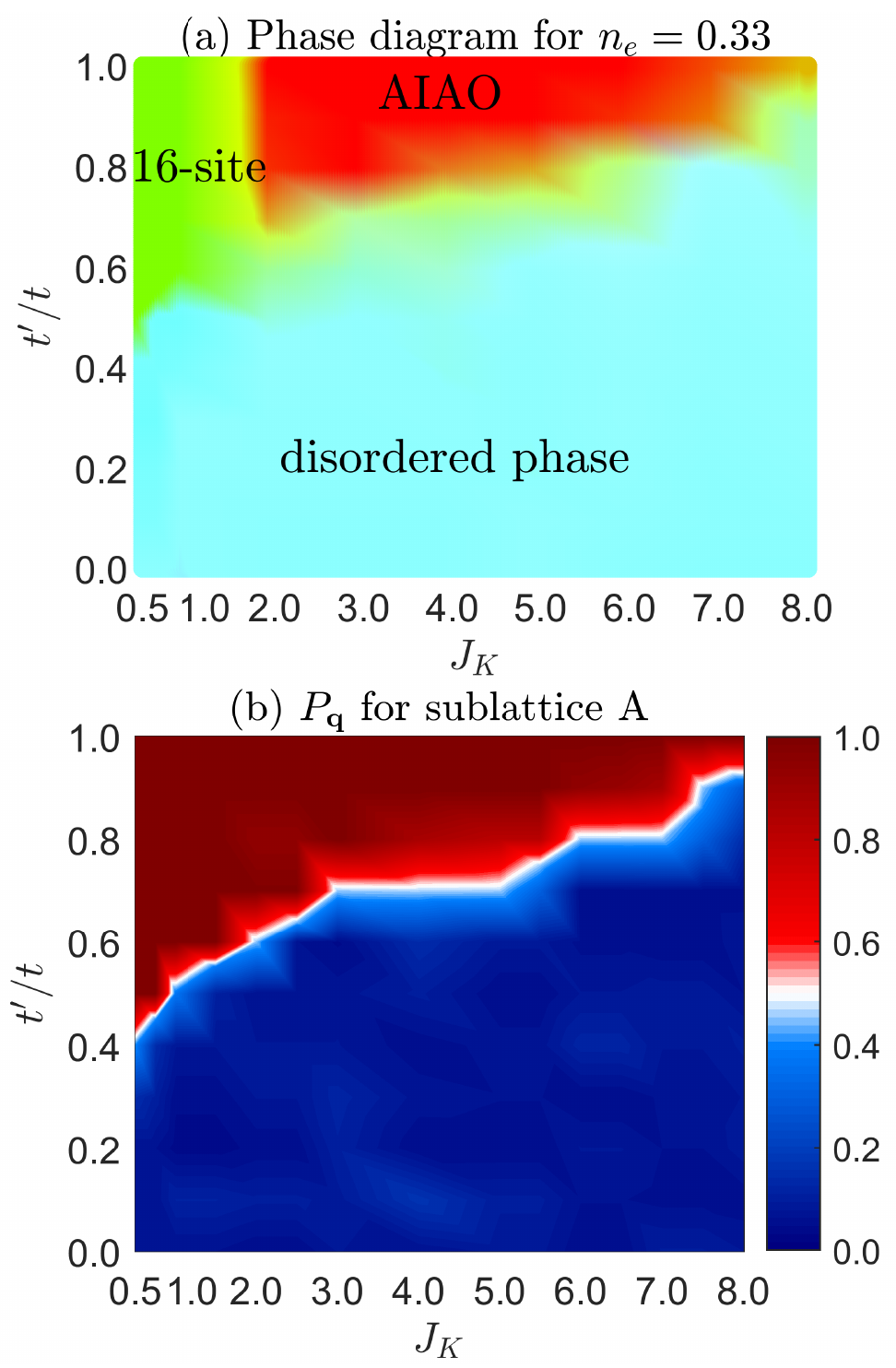}
	\caption{(Color online) (a) The phase diagram as a function of $t'/t$ and $J_K$ for number density of itinerant electron $n_e=0.33$. Again, a RGB scheme is used to represent different local correlation fractions. 
	(b) The order parameter $P_\mathbf{q}$ for sublattice $A$ plotted against $t'/t$ and $J_K$ for $n_e=0.33$ depicting the regions of the diagram with ordered and disordered phases.}
	\label{fig:phase_diag_mq_ne_0.33}
	\end{figure}
	
	\subsection{One-third filling}\label{subsec:one-third_filling}
	The phase diagram for $n_e=0.33$ is shown in Fig.~\ref{fig:phase_diag_mq_ne_0.33}(a) with $t'/t$ and $J_K$ as parameters. Here, we identify two ordered phases, an AIAO phase at large $J_K$ coupling, and a phase for which the magnetic unit cell consists of 16 sites at the small coupling limit. The hopping ratio $t'/t$ for which these phases are stabilized increases with the decrease of the Kondo coupling. 
		When the hopping ratio is small there is a disordered phase similar to the one found in the $n_e=0.50$ phase diagram. The difference between the ordered and disordered phases can also be seen in the order parameter $P_\mathbf{q}$, shown in  Fig.~\ref{fig:phase_diag_mq_ne_0.33}(b). The peak in spin structure factor for the AIAO phase appears at wave vector $\mathbf{q}=(0,0,0)$
	(as discussed in Section~\ref{subsec:half_filling}) and for the $16$-site phase the peak is observed at $\mathbf{q}=(\pi,\pi,-\pi)$ for sublattices $A, B$ and $C$
	and at $\mathbf{q}=(\pi,-\pi,\pi)$ for sublattice $D$.
	Taken separately, these two $\mathbf{q}$-vectors each imply a two-tetrahedron structure; combining them yields a four-tetrahedron or 16-site structure. 
	
\begin{figure}[tb]
	\centering
	\includegraphics[clip,trim=0cm 0cm 0cm 0cm,width=0.49\textwidth]{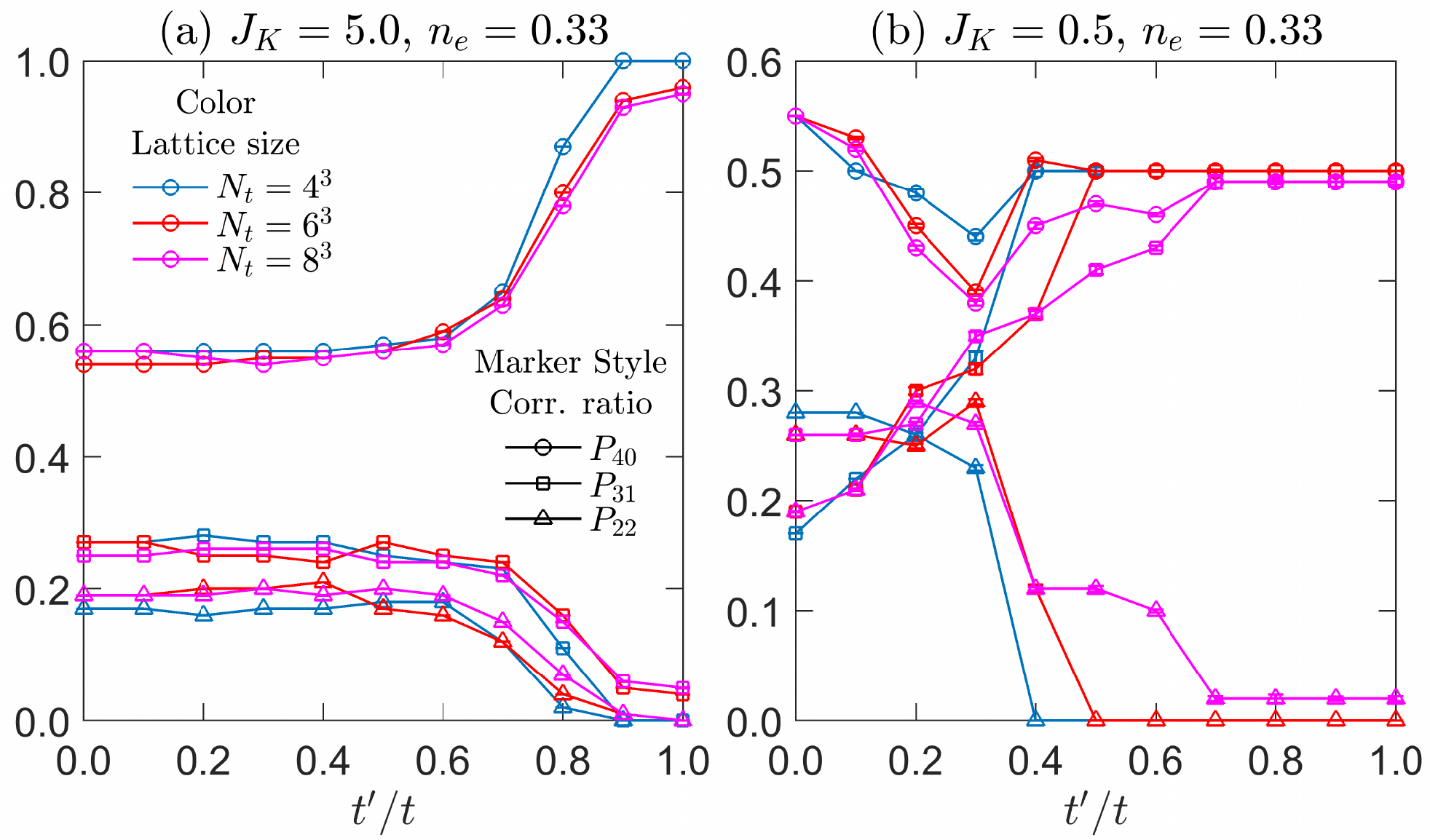}
	\caption{(Color online) (a) Local correlation fractions $P_{40}$, $P_{31}$ and $P_{22}$ as a function of $t'/t$ (a) for $J_K=5.0$ (AIAO state) and (b) for $J_K=0.5$ ($16$-site state) for three different system sizes.}
	\label{fig:line_plots_ne_0.33}
	\end{figure}

	In Fig.~\ref{fig:line_plots_ne_0.33}(a) the variation of local correlation fractions is shown as a function of $t'/t$  for $J_K=5.0$. The local spin configurations are AIAO ($P_{40}$) only at the isotropic limit ($t'/t\approx1$) and change to disordered configurations at intermediate and small values of $t'/t$. The spin configurations (not shown here) 
	change as $t'/t$ is reduced from an AIAO ordered state
	to all-in or all-out configurations on down-pointing tetrahedra and a combination of all spin configurations on up-pointing tetrahedra,
	similar to the $n_e=0.50$ case.
	Fig.~\ref{fig:line_plots_ne_0.33}(b) is for $J_K=0.5$, where the $16$-site phase occurs. In this phase, the spin configurations on half of the tetrahedra are all-in or all-out and on the other half are 3-in-1out. This statement is true for spin configurations on both down- and up-pointing tetrahedra, but below $t'/t\approx 0.4$ the down-pointing tetrahedra configurations become all-in and all-out while the up-pointing tetrahedra change to a disordered combination of all configurations.
	
	\subsection{One-quarter filling}\label{subsec:quarter_filling}
\begin{figure} [tb]
	\centering
	\includegraphics[clip,trim=0cm 0cm 0cm 0cm,width=0.40\textwidth]{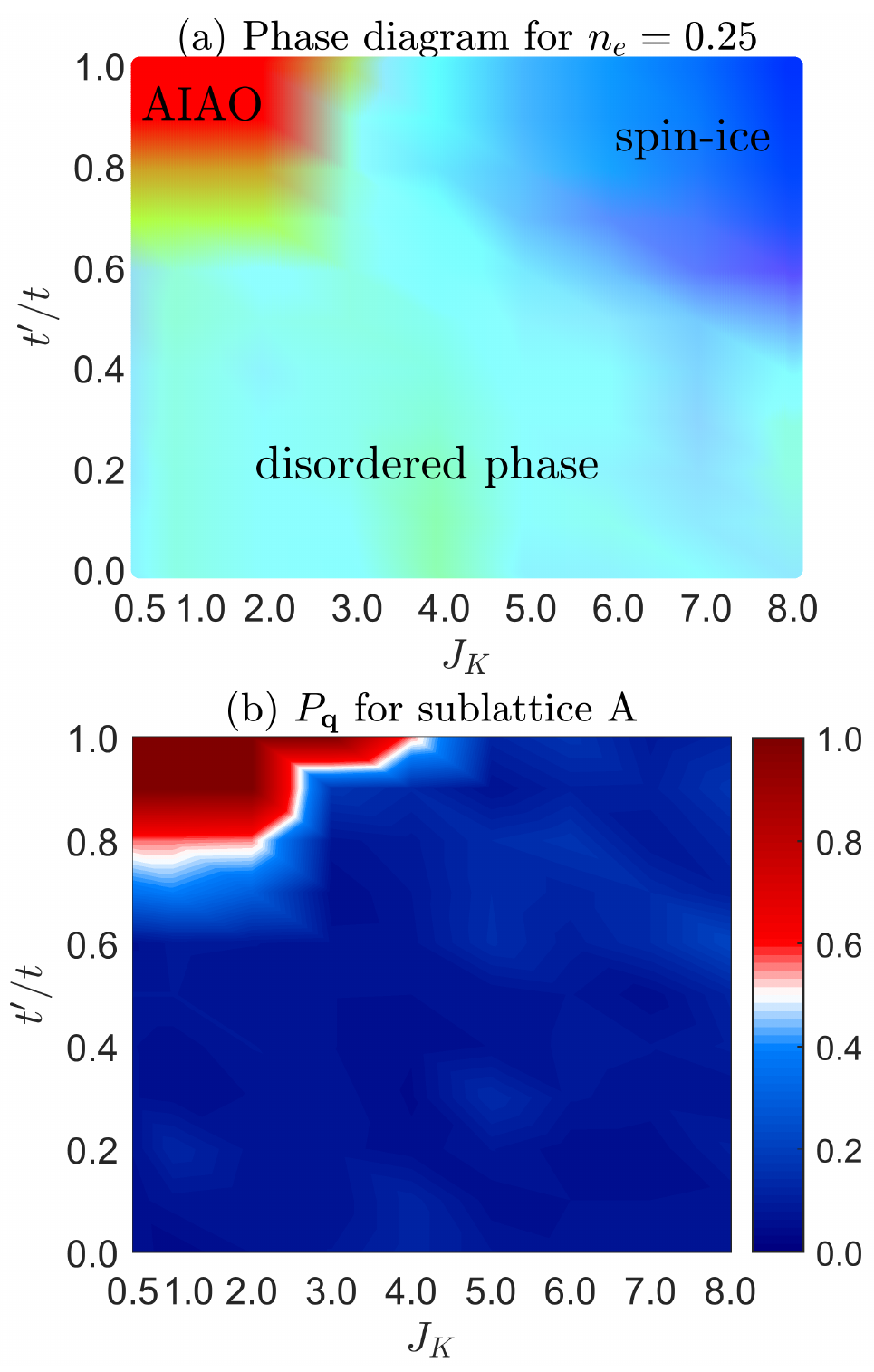}
	\caption{(Color online) (a) The phase diagram of localized spins as a function of Kondo coupling $J_K$ and hopping ratio $t'/t$ at one-quarter filling of itinerant electrons. Here also, a RGB scheme is used to represent different local correlation fractions. 
	(b) The order parameter $P_\mathbf{q}$ for sublattice $A$ plotted against $t'/t$ and $J_K$ for $n_e=0.25$. The regions of the diagram with ordered and disordered phases are clearly distinguishable.}
	\label{fig:phase_diag_mq_ne_0.25}
	\end{figure}

	Next, we discuss the phase diagram at one-quarter filling of itinerant electrons, shown in Fig.~\ref{fig:phase_diag_mq_ne_0.25}(a). In the upper half of the diagram there are two phases, a SI at large $J_K$ coupling and an AIAO state at small Kondo coupling. At small values of hopping ratio a disordered phase is realized. In the isotropic limit, as mentioned earlier, the double exchange mechanism is responsible for magnetic ordering at large $J_K$. In this limit, the itinerant electrons are fully aligned in the direction of local spins at each site and hopping processes contribute substantially if the localized spins are parallel. That means that FM order will be likely to dominate over AFM order for all values of number densities of itinerant electrons except at half filling. On a pyrochlore lattice, the FM interactions are frustrating and yield a SI ground state for Ising spins where in each tetrahedron two spins are forced to point towards the center while the other two away from it. In the current model,
	at one-quarter filling, every tetrahedron retains the 
	locally ferromagnetic 2-in-2-out ice-rule configuration. As
	shown in Fig.~\ref{fig:phase_diag_mq_ne_0.25}(b),
	the peak in $S(\mathbf{q})$ is very small, indicating no or weak long-range order.
	At small $J_K$, the AIAO phase can be understood in terms of an effective RKKY Hamiltonian with AFM NN interactions, as discussed for $n_e=0.50$ case. This is an ordered phase with a peak in $S(\mathbf{q})$ at $\mathbf{q}=(0,0,0)$. For the disordered phase, at small and intermediate values of $t'/t$, there is no magnetic order, as shown in lower half of Fig.~\ref{fig:phase_diag_mq_ne_0.25}(b).
	
\begin{figure}[tb]
	\centering
	\includegraphics[clip,trim=0cm 0cm 0cm 0cm,width=0.49\textwidth]{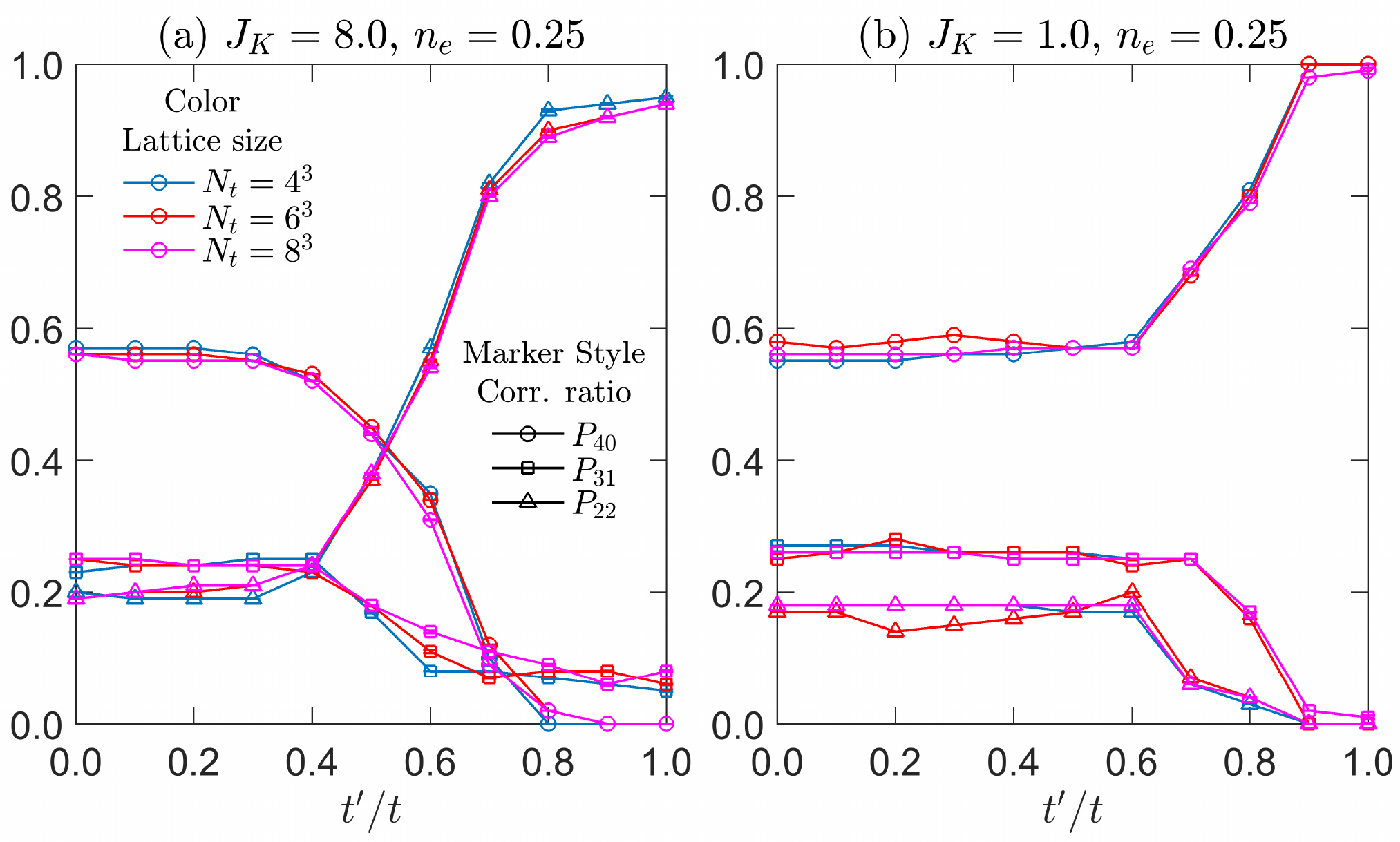}
	\caption{(Color online) (a) Local correlation fractions $P_{40}$, $P_{31}$ and $P_{22}$ as a function of $t'/t$ (a) for $J_K=8.0$ (SI state) and (b) for $J_K=1.0$ (AIAO state) 
	for three different system sizes at one-quarter filling.}
	\label{fig:line_plots_ne_0.25}
	\end{figure}

	Fig.~\ref{fig:line_plots_ne_0.25}(a) shows the local spin fractions vs.\ $t'/t$ for $J_K=8.0$ at 1/4 filling. In the isotropic limit, the spin configurations on most of the tetrahedra are 2-in-2-out. $P_{22}$ decreases as the hopping ratio $t'/t$ decreases, and a crossover is observed between $P_{40}$ and $P_{22}$. The spin configurations on both type of tetrahedra are 2-in-2out for $t'/t=1$, but change to a mixture of all-in and all-out states on down-pointing tetrahedra and to a disordered set of states on up-pointing tetrahedra when the hopping ratio is reduced. We show the variation of local correlation fractions as a function of $t'/t$ for $J_K=1.0$ 
	in Fig.~\ref{fig:line_plots_ne_0.25}(b). The AIAO type ordering becomes a disordered phase upon decreasing the hopping ratio. The spin configurations on down-pointing tetrahedra change from all-in or all-out to all-in and all-out while for up-pointing tetrahedra these change from all-in or all-out to a combination of all configurations.
	
	\subsection{One-sixth filling}\label{subsec:one-sixth_filling}
\begin{figure} [htb]
	\centering
	\includegraphics[clip,trim=0cm 0cm 0cm 0cm,width=0.40\textwidth]{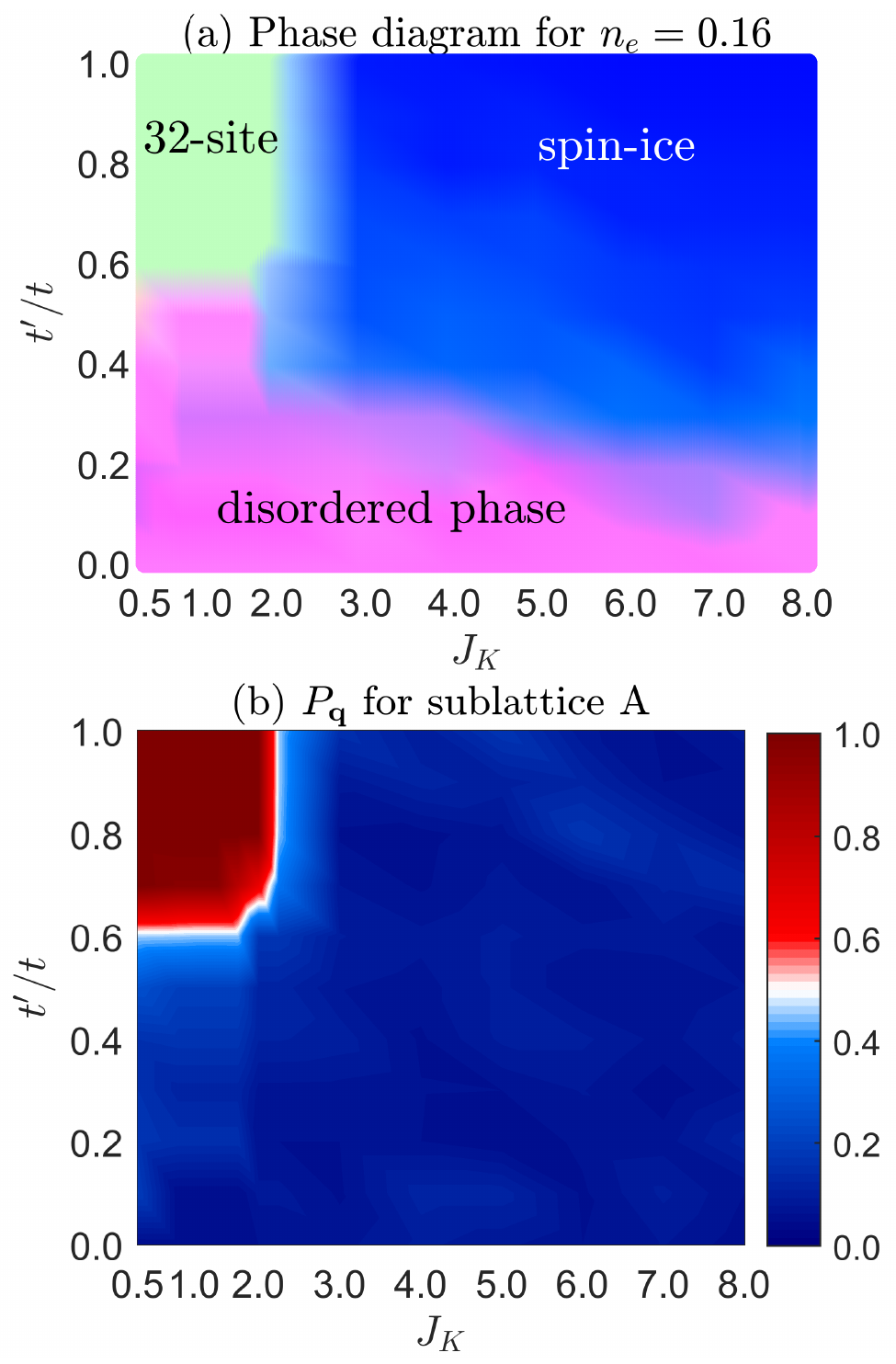}
	\caption{(Color online) (a) The phase diagram as a function of $t'/t$ and $J_K$ for number density of itinerant electrons $n_e=0.16$. Again, a RGB scheme is used to represent different local correlation fractions.
	(b) The order parameter $P_\mathbf{q}$ for sublattice $A$ plotted against $t'/t$ and $J_K$ for $n_e=0.16$. The regions of phase diagram with ordered and disordered phases are evident.}
	\label{fig:phase_diag_mq_ne_0.16}
	\end{figure}

	Finally, we discuss the magnetic phase diagram at one-sixth filling of itinerant electrons as shown in Fig.~\ref{fig:phase_diag_mq_ne_0.16}(a). In the isotropic limit, at large $J_K$ coupling, we observe a SI phase, while at small coupling the system develops an ordered phase whose unit cell consists of $32$ sites. In the large coupling limit, the double exchange mechanism governs the stabilization of the SI phase. This occurs due to the stabilization of FM ordering at this intermediate filling of itinerant electrons. The SI phase is a disordered phase, as can be seen in the plot of $P_\mathbf{q}$ in Fig.~\ref{fig:phase_diag_mq_ne_0.16}(b). For small $J_K$ coupling, the NN interactions in an effective RKKY Hamiltonian for the number density under consideration 
	are irrelevant and next-nearest neighbor interactions are AFM, which stabilize the complicated $32$-site phase. This phase is an ordered phase with peaks in $S(\mathbf{q})$ at $(\pi,\pi,\pi)$, $(-\pi,\pi,\pi)$, $(\pi,-\pi,\pi)$ and $(\pi,\pi,-\pi)$ for sublattices $A, B, C$ and $D$ respectively. Considering these
	$\mathbf{q}$-vectors together, the magnetic structure is found to be periodic over two tetrahedra in three directions, resulting in a $8$-tetrahedron or $32$-site phase. 
	In this magnetic structure, the spin configurations along 
	a particular direction on the pyrochlore lattice have a ``in-in-out-out'' ordering i.e., all the next-nearest neighbor
	spins are AFM. 
	The spin configurations of one half of the tetrahedra are 3-in-1-out, while one-sixth of them are 2-in-2-out and further one-eighth are all-in or all-out, which is a combination of all possible spin configurations on a tetrahedron. 
	
	It is important to note here that although we obtain a qualitative picture of the phase diagram from an effective RKKY Hamiltonian, the true nature of the complicated phases such as the $32$-sites and $16$-sites ordering is hard to predict from a simple RKKY analysis. With the decrease of $t'/t$ ratio, both at large and small Kondo coupling, a disordered phase is realized. This disordered phase is different as the spin configurations on more than half of the tetrahedra are 2-in-2out as compared to other number densities where spin configurations on more than half of the tetrahedra are all-in or all-out. 
	
\begin{figure}[htb]
	\centering
	\includegraphics[clip,trim=0cm 0cm 0cm 0cm,width=0.49\textwidth]{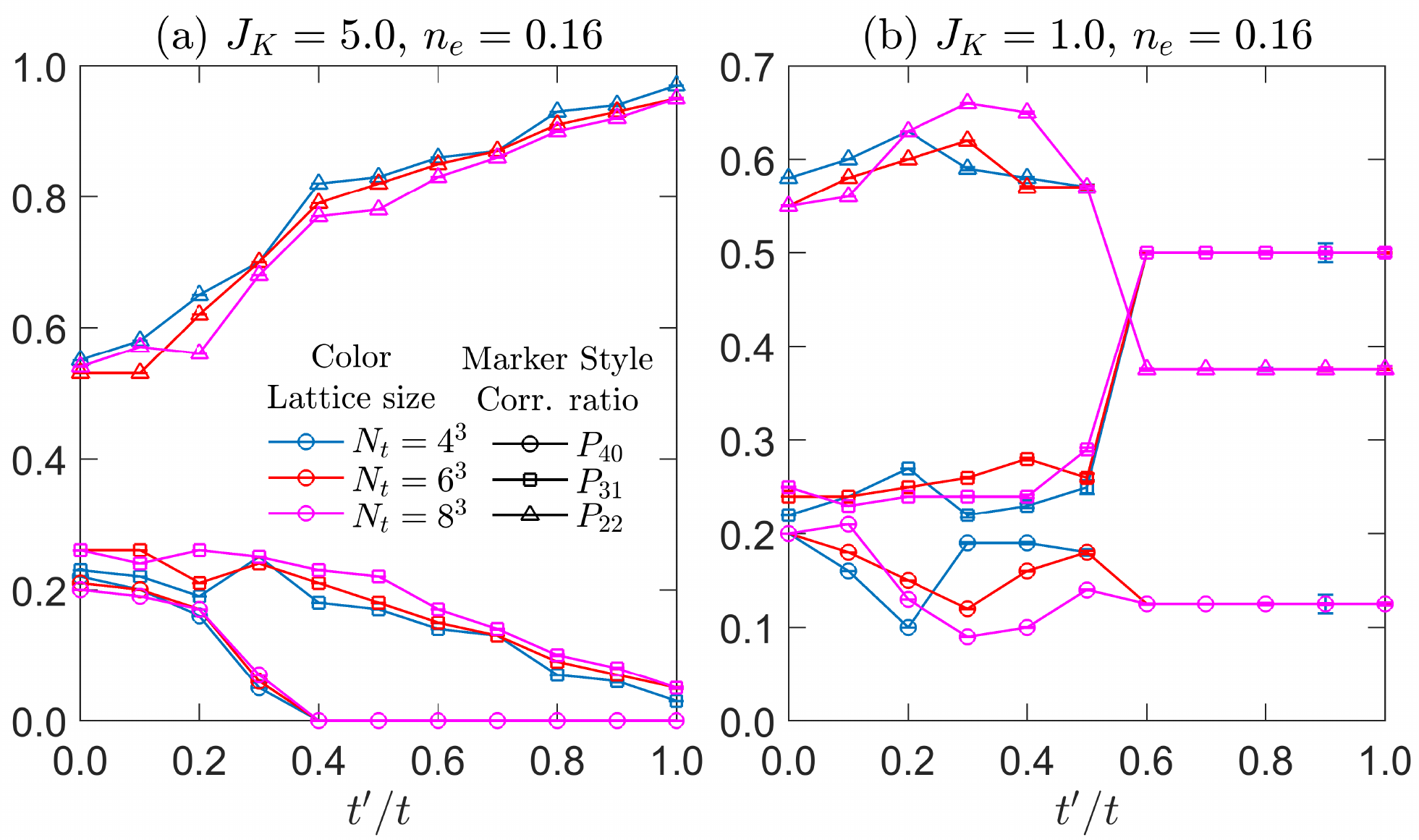}
	\caption{(Color online) Local correlation fractions $P_{40}$, $P_{31}$ and $P_{22}$ as a function of $t'/t$  for (a) $J_K=5.0$ (SI state) and (b) $J_K=1.0$ ($32$-site state) for three different system sizes.}
	\label{fig:line_plots_ne_0.16}
	\end{figure}

	We plot the local correlation fractions at two values of  $J_K$  in Fig.~\ref{fig:line_plots_ne_0.16} as a function of $t'/t$. For $J_K=5.0$, the spin configurations on most of the tetrahedra are 2-in-2-out for isotropic and intermediate values of $t'/t$, indicating a SI phase across this range. However, at small values of $t'/t$ a  reduction in $P_{22}$ is observed. In fact, the spin configurations on down-pointing tetrahedra are 2-in-2-out while on up-pointing tetrahedra the spin configurations are a combination of all configurations. For $J_K=1.0$ (Fig.~\ref{fig:line_plots_ne_0.16} (b)), in the isotropic limit, the spin configurations on half of the tetrahedra are 3-in-1-out, one-sixth are 2-in-2-out and one-eighth are all-in or all-out -- the spin configurations of the $32$-site ordered phase. At  intermediate values of $t'/t$, there is a crossover to a disordered phase. The spin configurations on down-pointing tetrahedra are all 2-in-2-out while on up-pointing tetrahedra there is a mixture of all configurations.
	
	\section{Summary}\label{sec:summary}
	We investigated a Kondo lattice model on a breathing pyrochlore lattice with strong easy-axis along the $\avg{111}$ direction. 
	A rich variety of ordered phases, including AIAO, SI, $16$-site and $32$-site orders, as well as a disordered phase, are stabilized due to competing effects of frustration, interactions with itinerant electrons, and frustration-relieving lattice distortion.
	
	\begin{acknowledgments}
	It is a pleasure to thank H. Ishizuka for helpful discussions. We acknowledge the use of GPU clusters at ACENET and Compute Canada for our numerical simulations. This work was supported by the Natural Sciences and Engineering Research Council of Canada (grant number 2020-05615). K.B. acknowledges support from the Center of Materials Theory as a part of the Computational Materials Science (CMS) program, funded by the U.S. Department of Energy, Office of Science, Basic Energy Sciences, Materials Sciences and Engineering Division.
	\end{acknowledgments}
	\appendix
	
	\section{Kernel Polynomial Method}\label{sec:KPM}
	
	In this section we review the KPM largely following the presentation in Ref.~\onlinecite{Weisse06}. Given an unscaled Hamiltonian $H_{0}$ with units of energy, one can define
    \begin{equation}
    H=\frac{H_{0}-\epsilon_{\mathrm{min}}}{\epsilon_{\mathrm{max}}-\epsilon_{\mathrm{min}}}-I,
    \end{equation}
    such that all eigenvalues of $H$ have magnitude less than 1. To find
    approximate bounds $\epsilon_{\mathrm{min}}$ and $\epsilon_{\mathrm{max}}$
    on the extreme eigenvalues of $H_{0}$, one can employ, e.g., the
    Lanczos method.

    The scaled matrix $H$ is a convenient starting point for performing a
    Chebyshev polynomial expansion. The Chebyshev polynomials satisfy
    $T_{m}(x)=\mathrm{cos}(m\arccos x)$ for $|x| \leq 1$. Via this identity,
    one can establish a close relationship between Chebyshev and Fourier
    cosine series.

    For an arbitrary function $F$, one can approximate
    \begin{equation}
    F(x)\approx\sum_{m=0}^{M-1}c_{m}T_{m}(x),\label{eq:F_expand}
    \end{equation}
     which is valid when $|x| \leq 1$. The coefficients
    \begin{equation}
    c_{m}=\frac{1}{\pi}(2-\delta_{0,m})g_{m}^{M}\int_{-1}^{+1}\frac{T_{m}(x)F(x)}{\sqrt{1-x^{2}}}dx
    \end{equation}
    can be accurately evaluated using Chebyshev-Gauss
    quadrature. Equality in Eq.~(\ref{eq:F_expand}) would be exact in
    the limit $M\rightarrow\infty$ and $g_{m}^{\infty}=1$. At finite
    truncation order $M$ it is useful to employ damping coefficients
    \begin{equation}
    g_{m}^{M}=\frac{(M-m+1)\cos\frac{\pi m}{M+1}+\sin\frac{\pi m}{M+1}\cot\frac{\pi}{M+1}}{M+1}
    \end{equation}
    corresponding to the Jackson kernel~\cite{Jackson12, Weisse06}. In a certain sense, these coefficients
    optimally damp artificial oscillations due to the Gibbs phenomenon.

    The Chebyshev polynomial expansion also works for matrices
    \begin{equation}
    F(H)\approx\sum_{m=0}^{M-1}c_{m}T_{m}(H).\label{eq:F_expand_H}
    \end{equation}
    To verify this, one can consider $H$ in its diagonal basis, and apply
    Eq.~(\ref{eq:F_expand}) to each eigenvalue separately.

    Chebyshev polynomials satisfy a numerically stable two-term recurrence,
    \begin{equation}
    T_{m}(H)=\begin{cases}
    I & \mathrm{if}\,m=0\\
    H & \mathrm{if}\,m=1\\
    2HT_{m-1}(H)-T_{m-2}(H) & \mathrm{if}\,m\geq2.
    \end{cases}\label{eq:recurr}
    \end{equation}
    That is, one can iteratively calculate each $T_m(H)$ from previous ones. The most numerically expensive part of each iteration is multiplying the matrices  $H$ and $T_{m-1}(H)$. The matrix dimensions of $H$ and $T_m(H)$ are proportional to system size $N$. Typically $H$ will be sparse, so that each matrix multiplication costs $\mathcal O(N^2)$ operations. The total cost to approximate $F(H)$ in Eq.~\eqref{eq:F_expand_H} then scales like $\mathcal O(M N^2)$.

    One can achieve a cost that scales \emph{linearly} in system size $N$ through stochastic approximation. The trace of $F(H)$ may be approximated as
    \begin{equation}
    \mathrm{Tr}\,F\approx\mathrm{Tr}\,R^{\dagger}FR,\label{eq:stoch_approx}
    \end{equation}
    where $R$ is a suitable random matrix with, typically, $N_{R} \ll N$ columns. More columns $N_{R}$ increases the computational cost but reduces the stochastic error, $\mathrm{Tr}\,(RR^{\dagger}-I)F$.
    The approximation is unbiased if $\langle RR^{\dagger}\rangle=I$.
    This is satisfied, for example, by independently drawing matrix
    elements $R_{ij}$ from a Gaussian distribution with standard deviation
    $N_{R}^{-1/2}$. In that case the stochastic error in Eq.~(\ref{eq:stoch_approx})
    would decay like $N_{R}^{-1/2}$. One can improve this scaling of error by using probing methods that take advantage of the decay typically present in matrix elements $F(H)_{ij}$~\cite{Tang12, Wang18b}.

    Combining the approximations of Eqs.~(\ref{eq:F_expand_H}) and~(\ref{eq:stoch_approx})
    yields
    \begin{equation}
    \mathrm{Tr}\,F(x)\approx\sum_{m=0}^{M-1}c_{m}R^{\dagger}\alpha_{m},\label{eq:tr_F_expand}
    \end{equation}
    where $\alpha_{m}=T_{m}(H)R$. Using Eq.~(\ref{eq:recurr}) one arrives
    at
    \begin{equation}
    \alpha_{m}=\begin{cases}
    R & \mathrm{if}\,m=0\\
    HR & \mathrm{if}\,m=1\\
    2H\alpha_{m-1}-\alpha_{m-2} & \mathrm{if}\,m\geq2.
    \end{cases}
    \end{equation}
    Again assuming sparsity of $H$, each matrix multiplication now
    costs $\mathcal{O}(N_{R}N)$ operations. The total computational cost
    to estimate $\mathrm{Tr}\,F(x)$ using stochastic approximation then scales as $\mathcal{O}(MN_{R}N)$, i.e., linear in system size $N$.
	
	\bibliographystyle{apsrev4-1}
	\bibliography{pyrochlore_project}
    \end{document}